\documentclass[prb,reprint,floatfix,superscriptaddress]{revtex4-1}
\usepackage{physics}
\usepackage{siunitx}
\DeclareSIUnit\BohrMagneton{\ensuremath{\mu_{\text{B}}}}
\DeclareSIUnit\formulaunit{f.u.}
\DeclareSIUnit\atomicunit{a.u.}
\DeclareSIUnit\arbunit{arb.unit}
\DeclareSIUnit\torr{Torr}
\DeclareSIUnit\rpm{RPM}

\usepackage{booktabs}

\usepackage{tikz}
\usepackage{tikz-3dplot}
\usetikzlibrary{intersections,calc,decorations.markings,decorations.pathmorphing,positioning}

\definecolor{gdviolet}{rgb}{0.282,0.204,0.537}
\definecolor{gdred}{rgb}{0.847,0.157,0.106}
\definecolor{gdorange}{rgb}{0.937,0.620,0.004}
\definecolor{gdgreen}{rgb}{0.361,0.675,0.145}
\definecolor{gdcyan}{rgb}{0.008,0.714,0.675}
\definecolor{gdblue}{rgb}{0.000,0.376,0.675}
\definecolor{gdyellow}{rgb}{0.808,0.824,0.004}

\newcommand{\mrx}{\mbox{Mn$_2$Ru$_x$Ga}}
\newcommand{\mrg}{\mbox{MRG}}
\newcommand{\tcmp}{T_{\text{comp}}}

\begin{document}
\title{Giant spin-orbit torque in a single ferrimagnetic metal layer}
\author{Simon Lenne}
\author{Yong-Chang Lau}
\affiliation{CRANN, AMBER and School of Physics, Trinity College Dublin,
Ireland}
\author{Ajay Jha}
\author{Gwenaël Y. P. Atcheson}
\affiliation{CRANN, AMBER and School of Physics, Trinity College Dublin,
Ireland}
\author{Roberto E. Troncoso}
\author{Arne Brataas}
\affiliation{Center for Quantum Spintronics, Department of Physics, Norwegian
University of Science and Technology, NO-7491 Trondheim, Norway}
\author{J.M.D.\,Coey}
\author{Plamen Stamenov}
\author{Karsten Rode}
\email[Corresponding author:]{rodek@tcd.ie}
\affiliation{CRANN, AMBER and School of Physics, Trinity College Dublin,
Ireland}

\begin{abstract}
Antiferromagnets and compensated ferrimagnets offer opportunities to
investigate spin dynamics in the `terahertz gap' because their resonance modes
lie in the \SIrange{0.3}{3}{\tera\hertz} range. Despite some inherent
advantages when compared to ferromagnets, these materials have not been
extensively studied due to difficulties in exciting and detecting the
high-frequency spin dynamics, especially in thin films. Here we show that
spin-orbit torque in a single layer of the highly spin-polarized compensated
ferrimagnet \mrx\ is remarkably efficient at generating spin-orbit fields
$\mu_0 H_{\text{eff}}$, which approach
\SI[per-mode = symbol]{0.1e-10}{\tesla\per\ampere\metre\squared} in the
low-current density limit -- almost a thousand times the Oersted field, and one to
two orders of magnitude greater than the effective fields in heavy metal/ferromagnet
bilayers.
\end{abstract}

\maketitle

We depend on fast, reliable exchange of information across
long distances through intercontinental optical fibres, as well as 
short-distance connections between the central processing unit of a computer and its memory. The latter is
the bottleneck to the powerful computing
facilities needed
in a future where
machine learning and algorithms aid our daily lives. This bottleneck
is difficult to overcome because electronics lack a practical
chip-based solution to produce and detect electromagnetic waves in the spectral
range between \SIlist{0.3; 30}{\tera\hertz} known as the terahertz gap.

\citet{PhysRevB.39.6995} realised that angular momentum
could be transferred from one magnetic layer (a polariser) to another (the analyser) by a spin polarised current.\cite{SLONCZEWSKI1996L1} 
This spin-transfer
torque has enabled the scaling of devices that depend on the relative magnetic
orientation of two ferromagnetic layers.\cite{Chappert2007}

Spin electronics exploiting the orbital degree of freedom of the electron
is a recent development. A major advance was the discovery
that the angular momentum could be
supplied by a diffusive spin current\cite{miron2011perpendicular,LiuScience2012_SHE_Ta} created via the spin Hall
effect\cite{d1971possibility,PhysRevLett.83.1834} in a non-magnetic
heavy metal layer adjacent to the ferromagnet.
Devices based on these bilayers require a bare minimum of two
layers on a substrate. Earlier,
\citet{PhysRev.100.580} and \citet{bychkov1984properties} had shown that in
crystalline or patterned structures lacking inversion symmetry, a
current-induced spin polarisation (CISP) is a direct consequence of the
symmetry of the band structure. This idea was developed by
\citet{PhysRevLett.113.157201}\cite{zelezny2017}
to predict the form of the tensor relating the charge current to the
CISP in crystals of different symmetry. \SI{90}{\degree} switching of the metallic antiferromagnets
CuMnAs\cite{wadley2016} and Mn$_2$Au\cite{Mn2Au2018} was subsequently observed. These
ground-breaking results established the existence of a current-induced,
field-like (or reactive) torque, and allow an estimate of the strength of the effective magnetic field
by comparing it to the in-plane magnetic anisotropy of the
material.

Hitherto there has been no quantitative measurement of the anti-damping (or
dissipative) spin-orbit torque in homogeneously magnetised ferrimagnetic or
antiferromagnetic single-layer
samples. Here we show, via harmonic analysis of the anomalous Hall effect,\cite{PhysRevB.89.144425}
that in a single layer of the prototype
half-metallic compensated ferrimagnet \mrx\ with $x=0.7$
(\mrg),\cite{KurtPRL2014,Thiyagarajah2015,ZicPRBRapid2016,BorisovAPL2016,BettoAIPAdvances2016,Betto2015,doi:10.1063/1.5001172}
both the field- and damping-like components of the torque reach record values,
almost two orders of magnitude stronger than those obtained in the
bilayer ferromagnet/heavy metal systems, or in metallic
ferromagnets\cite{Ciccarelli2016} and semimagnetic semiconductors\cite{Kurebayashi2014}.
The record values of the single-layer SOT and the dominance of the dissipative
torque, open a path to sustaining magnetic oscillations in the
terahertz gap.

\begin{figure}
  \centering
  \includegraphics[width=\columnwidth]{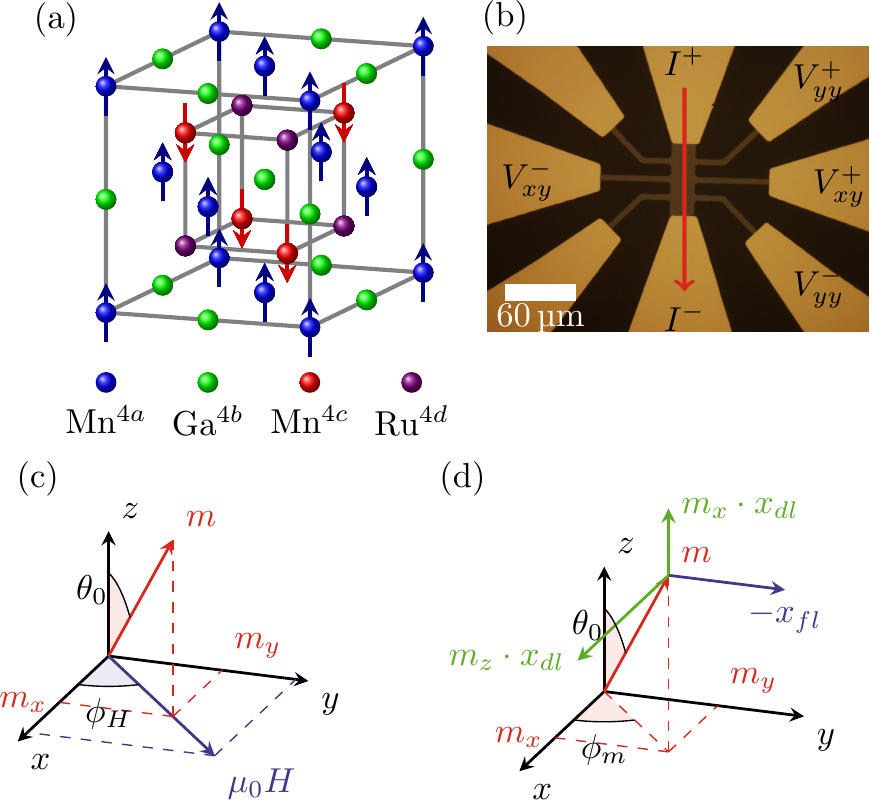}
  \caption{(a) \mrg\ crystal structure. The current is carried mainly by
    electrons in bands originating from Mn in the $4c$ position, which has point
    group symmetry $\bar{4}3m$. Arrows show the direction of the magnetic
    moment on each site. (b) Micrograph of a Hall
    bar with the contacts labelled.
    (c) Illustration of the coordinate system: $\theta_0$ is the
    polar angle of the magnetisation vector in the absence of the SOT field,
    and $\phi_m = \phi_H$ is the azimuthal angle of the magnetisation and applied
    field vectors, respectively. (d) Illustration of the
    effective SOT fields acting on the magnetisation with a bias current along
    \mrg\ $\left[ 010 \right] \parallel y$, the field-like (reactive)
    component in blue, and the two 
    damping-like (dissipative) components in green.}%
  \label{fig:intro}
\end{figure}
Thin-film samples of \mrg\ grown on MgO by DC-magnetron sputtering from
stoichiometric targets crystallise in a Heusler-like structure, space
group $F\bar{4}3m$ illustrated in  \figurename~\ref{fig:intro}a, where the conduction bands originate predominantly from Mn in 
$4c$ sites.\cite{Betto2015} The films are patterned into the micron-sized Hall bar
structures shown in \figurename~\ref{fig:intro}b, where the bias current $j$ is
parallel to the \mrg\ $\left[ 010 \right]$ axis. Further details on sample
growth can be found in the supplementary material and
[\onlinecite{BettoAIPAdvances2016}].  We determine
the current-induced effective fields via the anomalous Hall effect (AHE),
assuming it is proportional to the $z$ component of the
magnetisation of the Mn$^{4c}$ sublattice. Due to the substrate-induced
biaxial strain, the point group of Mn in this position is reduced from $\bar{4}3m$ to
$\bar{4}2m$. Here we restrict our analysis to the effect on one sublattice, as
the other will follow via inter-sublattice exchange with a phase-lag.
We treat all effective
torques as equivalent to external applied fields. For an
in-plane applied field $H$, the magnetisation is described by the
polar and azimuthal angles $\theta_0$ and $\phi_m$, with the latter 
taken to be equal to the azimuthal angle $\phi_H$ of the applied field
because the four-fold in-plane anisotropy is weak compared with the
uniaxial perpendicular anisotropy. The coordinates describing the
magnetic state are shown in \figurename~\ref{fig:intro}c. In the presence of a
unit charge current density $j \parallel \left[ 010 \right]$,
the CISP produces a SOT effective field (see
\figurename~\ref{fig:intro}d):
\begin{equation}
  \mu_0 \mathbf{h_{\text{SOT}}} = m_z x_{dl} \mathbf{e_x} - x_{fl} \mathbf{e_y} + m_x x_{dl}
  \mathbf{e_z}
  \label{eq:sot010}
\end{equation}
where $\mathbf{e_i}$ are unit vectors, $m_i$ are the components of the unit
magnetisation vector, and $x_{fl}$, $x_{dl}$ are the coefficients of the field-
and damping-like contributions to the spin-orbit field, respectively. The units of $\mu_0
\mathbf{h_{\text{SOT}}}$, $x_{fl}$
and $x_{dl}$ are then \si{\tesla\per\ampere\metre\squared}. Henry is an
equivalent unit.
When the bias
current has an alternating component ($j = j_{\text{dc}} + j_{\text{ac}}
\sin{\omega t}$) we detect the effect of the CISP on both the second and the third
harmonic response using lock-in demodulation. The conversion from the voltages detected
at the different harmonics to the magnitude of the effective fields is detailed
in the supplementary material. In \tablename~\ref{tab:harmonics} we indicate the
symmetry of the different contributions to $V_{xy}$ in the first, second and
third harmonic responses. Contributions from the anomalous Nernst effect (ANE) are suppressed by
measuring $V_{xy}^{3 \omega}$ or by taking the difference of $V_{xy}^{2
\omega}$ measured with positive and negative DC bias. The contribution from the homogeneous
temperature variation $\Delta T$ oscillating at twice the applied frequency is
determined from
data in \figurename~\ref{fig:sigmas}, as explained in the supplementary
material.

\begin{table}
  \caption{\label{tab:harmonics} Linear contributions to $V_{xy}$ 
    up to third order in current density. - means no contribution,
    \textsf{/} a contribution odd in  $j_{\text{dc}}$, \textsf{v} even in
    $j_{\text{dc}}$ and \textsf{o} independent
of current direction. $\sigma_{xy}$ contributes implicitly to
all four effects.}
  \addtolength{\tabcolsep}{3pt}
  \begin{tabular}{l c c c}
    \toprule[1pt]
    Contribution/Harmonic         & $\omega$ & $2 \omega$ & $3 \omega$ \\
    \midrule[0.5pt]
    \addlinespace[0.5pt]
    Anomalous Hall Effect: $\sigma_{xy}$                 & \textsf{o} & - & - \\
    Anomalous Nernst Effect: $\partial T/\partial z$           & - & \textsf{v} & - \\
    Homogeneous $\Delta T$ oscillating at $2 \omega$: $\partial \sigma / \partial T$    & \textsf{v} & \textsf{/} & \textsf{o} \\
    Current-induced fields: $\mathbf{h_{\text{SOT}}}$              & \textsf{/} \textsf{v} & \textsf{/} & \textsf{o} \\
    \bottomrule[1pt]
  \end{tabular}
\end{table}

\begin{figure}
  \centering
  \includegraphics[width=\columnwidth]{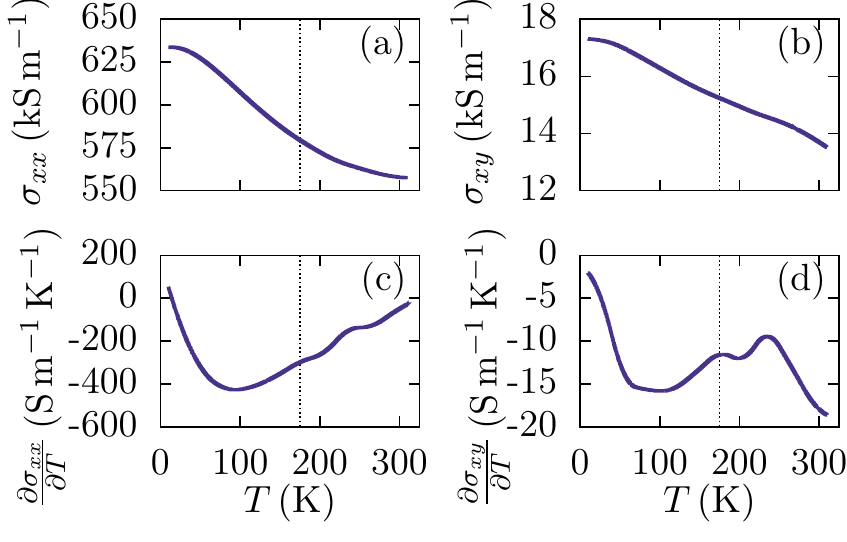}
  \caption{Temperature dependence of the longitudinal and transverse
    conductivity of \mrg~(a and b), and their temperature derivatives (c
    and d). The variation of $\sigma_{xy}$ follows the variation of
    the magnetisation of the $4c$ sublattice.\cite{Betto2015} The
    compensation temperature $\tcmp$ where the net magnetisation
    changes sign is \SI{175}{\kelvin} (vertical dashed line). Since the data were
    recorded at remnance, the direction of the sublattice moments does
    not change at $\tcmp$.}%
  \label{fig:sigmas}
\end{figure}
\figurename~\ref{fig:sigmas} shows the temperature dependence of the
longitudinal and transverse conductivity of \mrg , recorded
in the remnant state after saturation in a positive field at room temperature.
The conductivity $\sigma_{xx}$ (\figurename~\ref{fig:sigmas}a) increases with
decreasing $T$, and its saturation value of \SI{630}{\kilo\siemens\per\metre}
or $\left [ \SI{159}{\micro\ohm\centi\metre} \right ]^{-1}$ corresponds to the minimum metallic
conductivity of a bad metal where the mean free path is comparable to the
interatomic spacing.\cite{mott1990metal} The Hall conductivity $\sigma_{xy}$
(\figurename~\ref{fig:sigmas}b) closely follows the Mn$^{4c}$ sublattice
magnetisation.\cite{PhysRevB.98.220406}
The lower panels show the temperature-derivatives of $\sigma$.

\begin{figure}
  \centering
  \includegraphics[width=\columnwidth]{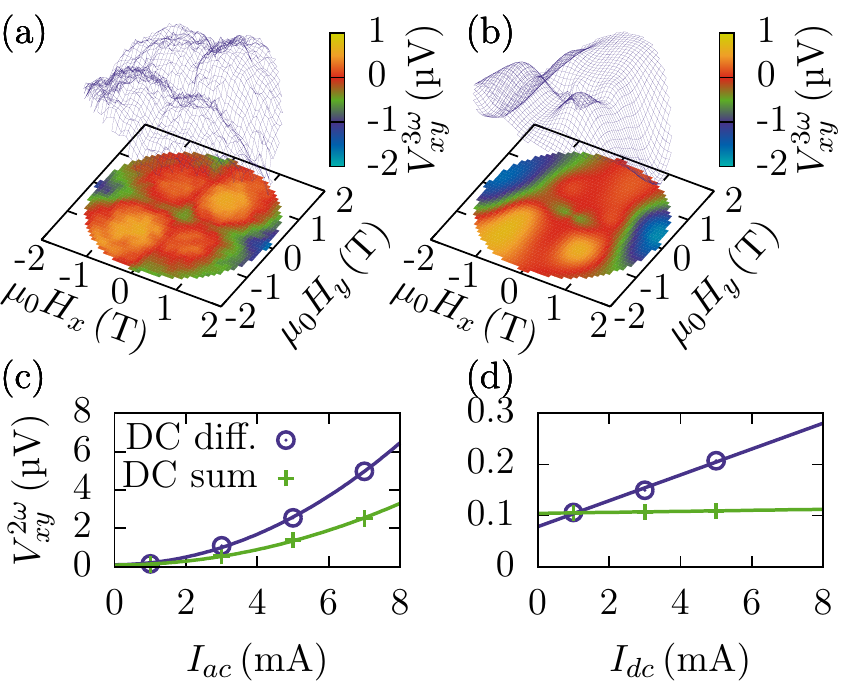}
  \caption{(a) Surface plot and its 2D colour map projection of the experimentally observed voltage
    at the \emph{third} harmonic $V_{xy}^{3\omega}$. (b) Calculated response based on the
    experimental values of $\cos{\theta_0}$.
    (c) and (d) show AC (with
    $I_{\text{dc}} = \SI{3}{\milli\ampere}$) and DC ($I_{\text{ac}} =
    \SI{1}{\milli\ampere}$) current-dependence of the \emph{second} harmonic
    $V_{xy}^{2 \omega}$ signal. By
    making the difference and sum of records made with positive and negative DC
    offset we isolate the SOT (the difference) from the anomalous Nernst effect ANE (the sum).
  }%
  \label{fig:mmsot}
\end{figure}
We now turn to the SOT\@. \figurename~\ref{fig:mmsot} a and b shows the
experimentally observed $V_{xy}^{3 \omega}$ and its calculated values based on the
experimental $\theta_0$. There is excellent
agreement between experiment and the model, which is based only on the site
symmetry, the data in \figurename~\ref{fig:sigmas}, and the first harmonic response (used to determine $\theta_0$ and the
anisotropy constants).
All the features in both the $\theta_0$ and
$\phi$ dependencies are well reproduced: two deep minima around $\mu_0
H_{x,\text{max}}$, four maxima that align with the four-fold in-plane
anisotropy due to the small value of the in-plane anisotropy constant $K_2^{'}$, as
well as a weaker central minimum at small fields. Qualitatively, the
field-dependence of the SOT can be understood by comparing equation~\eqref{eq:sot010} with 
\figurename~\ref{fig:mmsot} (a),
and noting that the magnitude of $\Delta \theta$ depends not only on $\theta_0$
itself, but explicitly on
the competition between the SOT field, the anisotropy field and the applied
field. At low applied fields ($\theta_0 \approx 0$) the current-driven wobble of $m$ is
determined by a combination of the anisotropy and SOT fields. $\Delta \theta$
is small, however, because $\cos{\theta_0} \approx 1$, hence the central  minimum is shallow. At higher applied fields, $\theta_0$ deviates from 0, but the SOT
field now has to compete with both the (higher) anisotropy field and the
Zeeman torque provided by the applied field acting on the
net magnetisation. This gives rise to the characteristic four-fold 
signal. An exceptional feature appears around $\mu_0 H_x \approx \pm \SI{2}{\tesla}$
where the damping-like field in the $z$ direction scaling as $m_x$ 
produces a field strong enough to dwarf both the anisotropy field and the applied
field.
We emphasise not only the qualitative
agreement, but also that the absolute magnitude of the signal
agrees very well with the model when we fit the coefficients
of the field- and damping-like fields; $x_{fl} =
\SI{-15e-13}{\tesla\per\ampere\metre\squared}$ and $x_{dl} =
\SI{50e-13}{\tesla\per\ampere\metre\squared}$.

We then
determine the dependence of the effective
field magnitude on bias current, (\figurename~\ref{fig:mmsot} c and d). A current of \SI{1}{\milli\ampere} in our
films is equivalent to $j$ of about $\SI{2.5e9}{\ampere\per\metre\squared}$.  We expect
$V_{xy}^{2 \omega}$ due to SOT to
scale with $I_{\text{dc}}$ and  $I_{\text{ac}}^2$, while the effects due to
thermal gradients should be independent of $I_{\text{dc}}$ and scale as
$I_{\text{ac}}^2$. Indeed, the DC difference is
quadratic in $I_{\text{ac}}$ (\figurename~\ref{fig:mmsot}c) and linear in
$I_{\text{dc}}$ (\figurename~\ref{fig:mmsot}d), while the DC sum is quadratic
in $I_{\text{ac}}$ and practically independent of $I_{\text{dc}}$.

It is instructive to compare the effective fields due to intrinsic
SOT with those recorded on conventional bilayers of a heavy metal
(typically Pt, Ta or W) and a $3d$ ferromagnet (typically Co, Fe, CoFe or
CoFeB).
For bilayers, the damping-like effective field per current
density can be written: $\mu_0 H_{\text{dl}} / j =
(\theta_{\text{SH}} \hbar) / (2 e M_s t)$, where $\theta_{\text{SH}}$ is the
spin-Hall angle of the heavy metal, $\hbar$ is the Planck's constant, $e$ is the
electron charge, $M_s$ the magnetisation of the ferromagnet and $t$ its thickness. For
\SI{1}{\nano\metre} of CoFeB ($M_s \approx \SI{1}{\mega\ampere\per\metre}$),
which has a magnetic moment equivalent to that of $\approx \SI{30}{\nano\metre}$ nearly compensated 
\mrg, and $\theta_{\text{SH}} = \SI{40}{\percent}$ we obtain an effective,
damping-like field of \SI{1.3e-13}{\tesla\per\ampere\metre\squared}
(\SI{0.13}{\pico\henry}). We
would need a fictitious spin-Hall angle of \SI{400}{\percent} to match the value
of the field-like term in \mrg\ and \SI{1200}{\percent} to match the
damping-like term. 

This comparison highlights the inherent advantage of using ferrimagnets
in combination with intrinsic SOT\@. In a bilayer, increasing the thickness of the ferromagnet
beyond the spin diffusion length (typically $< \SI{10}{\nano\metre}$), does
not produce any additional torque. 
If the ferromagnet is \SI{2}{\nano\metre} rather than \SI{1}{\nano\metre} thick, the effective
field may be reduced to half, whereas the field in single-layer \mrg\ is unchanged
with thickness. The volume of \mrg\
can be scaled up or down without changing the torque, providing the current
density is constant. The nature of the intrinsic torque is staggered acting directly on the 
Mn$^{4c}$  sublattice, hence a more correct comparison might be be to normalise the spin Hall angle
using the sublattice magnetisation, which is approximately ten times greater
than the net magnetisation at room temperature for the present sample.  Furthermore,
the torque is maintained even in the absence of any net magnetisation at
the ferrimagnetic compensation temperature, thus permitting GMR- and TMR-based
device structures to be excited by SOT even in the absence of any net
moment of the free layer. This enables targeted control of the dynamics,
and the excitation of both in-and out-of phase resonance modes.

\begin{figure}
  \centering
  \includegraphics{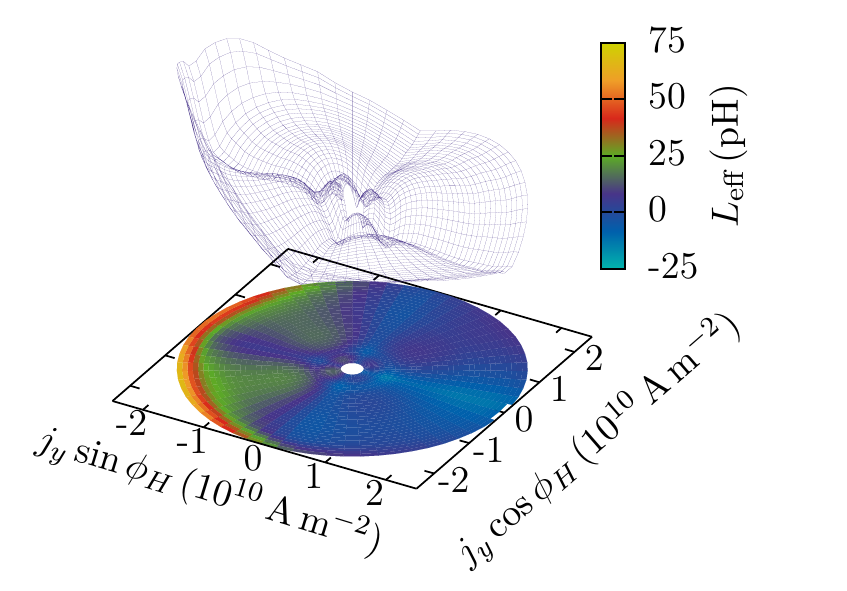}
  \caption{High-current-density effective induction $L$ in $\si{\pico\henry} =
    \si{\pico\tesla\metre\squared\per\ampere}$ as a function of the bias
    current density and the angle $\phi_H$ between $\mathbf{e_x}$ and the
    applied magnetic field $\mu_0 H_{\text{app}} = \SI{0.4}{\tesla}$. The
    effective inductance reaches $\sim \SI{75}{\pico\henry}$, corresponding to
    an effective field at $j=\SI{2.5e10}{\ampere\per\metre\squared}$ of $\mu_0
    H_{\text{eff}} \sim \SI{1.9}{\tesla}$. The interpretation of this striking
    result is  discussed in the main text.}
  \label{fig:high-j}
\end{figure}
The high effective fields found above assuming small, linear, current-driven
variations in $\theta_0$,
imply that the action of the SOT should also
be observable in the non-linear transfer characteristics of our Hall bar
device. We therefore proceeded as follows:

We first recorded a full field-in-plane hysteresis loop from
\SIrange{-14}{14}{\tesla} to determine the relation between 
$m_{z}$ and the applied field and invert
this relation numerically, to be able to deduce, from $m_{z}$, the value of the
total effective field at any given applied current. 
Then, a constant external
field $\mu_0 H = \SI{0.4}{\tesla}$ is applied in the sample plane and rotated
around $e_z$, changing the azimuthal angle $\phi_H$ from
\SIrange[range-phrase = {\text{ to }}]{0}{360}{\degree} while recording
$m_{z}$ (inferred from $V_{xy}$). This measurement is repeated for range of current densities from
\SIrange{0.2e10}{2.5e10}{\ampere\per\metre\squared}. As the action of the SOT
field depends directly on its direction relative to the direction of the
magnetisation ($\theta_M$ and $\phi_M \approx \phi_H$), we can
subtract any variation that is $\phi$-\emph{independent}. This $\phi$-independent effective field contains
all variations that are due to heating. The result, after subtraction, is shown in
\figurename~\ref{fig:high-j}, where we give the effective field in terms of the
effective, current-induced inductance in $\si{\pico\henry} =
\SI{1e12}{\tesla\metre\squared\per\ampere}$. A current density of
$j=\SI{2.5e10}{\ampere\per\metre\squared}$ can produce an effective inductance
$L_{\text{eff}} \approx \SI{75}{\pico\henry}$, equivalent to an effective in-plane field of
\SI{1.9}{\tesla}! We note that this field is sufficient to magnetically
switch $\approx \SI{2}{\percent}$ of the sample.

We make two important comments
on this analysis. First, by removing the $\phi$-independent part of the signal,
we also remove any SOT that behaves the same way. If we again assume the SOT
fields can be described by the tensors reported by \citet{zelezny2017} and
\citet{troncoso2018antiferromagnetic}, and
expand the relation between $h_{\text{eff}}$ and $\Delta \theta$ to second
order in $\Delta \theta$, we find that we have removed a damping-like contribution along $e_z$ that varies
as $m_x^2 + m_y^2$, which may be considerable. Second, as we are normalising
with respect to the action of the external, \emph{in-plane}, field, the
SOT effective field directed along $e_z$, remaining after the procedure
outlined above, contributes to our signal
$\propto {}^1\!/_{\cos{\theta_M}}$ as seen by the upturn at $\phi_H =
\SI{270}{\degree}$ in \figurename~\ref{fig:high-j}.


The strong effective SOT fields in MRG are related to its high anomalous Hall
angle.\cite{Thiyagarajah2015} The value is unusual in
the sense that \mrg\ does not contain any elements heavier than Ru; 
in any case the AHE angle does not scale with Ru content $x$. Furthermore the
conduction electrons in \mrg\ are predominantly $d$-like, although it has been suggested that Ga in the
Mn-containing Heuslers lends some $p$ character to the bands at the Fermi-level
through hybridisation, increasing the spin-orbit coupling of the
conduction electrons\cite{LauPRB2019}. We have already seen above that \mrg\ is at the limit of
metallic conductivity. From our measurements of $\sigma_{xx}$ and
$\sigma_{xy}$, we can deduce the spin-orbit scattering cross-section and find
that it corresponds to \SI{60}{\percent} of the unit cell surface area. The very
large scattering cross section is consistent with the very short mean free path.


So far we have demonstrated high current-induced effective fields as well as a high
ratio ($\sim 3$) of the dissipative (anti-damping) to the reactive (field-like)
torques. This will allow for the realization of more efficient magnetic
switching\cite{Garello2014}, exchange-bias manipulation\cite{Lin2019} as well
as low-current control of magnetic textures\cite{Hals2014}.
The key question is, whether sustained self oscillation can be driven by
SOT\@.
We address this from two different
angles, first by considering the results established by
\citet{troncoso2018antiferromagnetic}, noting that the effective fields will
act distinctly on the magnetisation and the Néel vectors. Using the numerical values of the
effective fields found in the linear, low-current regime, self-oscillations
will emerge for current
densities that provide a reactive torque which is sufficient to overcome the
in-plane anisotropy $\sim \SI{0.1}{\tesla}$ for \mrg, which corresponds to
$j \gtrsim \SI{7e10}{\ampere\per\metre\squared}$. The second necessary condition
is that the dissipative torque must overcome the Gilbert damping
$\alpha$. Taking $\alpha \approx 0.01$ we find the condition  
$j \gtrsim \SI{10e10}{\ampere\per\metre\squared}$. An alternative approach is to compare directly  
the effective inductance created by the SOT and the self inductance
of the oscillating element. In a shorted Hall bar device,
a crude estimate of the self inductance 
for a \SI{500}{\nano\metre} thick film with an active
length of \SI{20}{\micro\metre} is $\SI{0.1}{\pico\henry}$ -- the dimensions are chosen to enhance impedance
matching to free space in a real oscillator. We saw in \figurename~\ref{fig:high-j}
that the effective inductance reaches values two orders of magnitude
higher than this, ensuring that oscillatory behaviour is possible, even in the
low-current-density region. The natural
frequency of the oscillator will be determined by the larger of the two effective
inductances, that is by the SOT and the magnetic resonance frequency of the
material, which we 
previously estimated as $\SI{0.75}{\tera\hertz}$.\cite{fowley2018prb}


In summary, we find that current-induced spin orbit torque reaches record values in
single-layers of the compensated, half-metallic ferrimagnet \mrx, well in
excess of those achieved in bilayer structures. With realistic values of
damping, this should allow sustained magnetic oscillations that
could be detected by magnetoresistive effects,
or free-space emission using a suitable antenna.
A cheap, compact, and tunable oscillator operating in the terahertz gap
would break new ground in spin dynamics, and could potentially unlock a new realm
of information transfer at bandwidths three orders of
magnitude higher than those of the present day.

\begin{acknowledgments}
  This project has received funding from the \emph{European Union's} Horizon
  2020 research and innovation programme under grant agreement No 737038, and
  from Science Foundation Ireland through contracts 12/RC/2278 AMBER and
  16/IA/4534 ZEMS as well as the Research Council of Norway through its Centres
  of Excellence funding scheme, Project No. 262633 ``QuSpin''. The authors
  declare no competing financial interests.
\end{acknowledgments}

\bibliography{bibliography}

\end{document}


\title{Supplementary matriel to `Giant spin-orbit torque in a single ferrimagnetic metal layer'}
\author{Simon Lenne}
\author{Yong-Chang Lau}
\affiliation{CRANN, AMBER and School of Physics, Trinity College Dublin,
Ireland}
\author{Ajay Jha}
\author{Gwenaël Y. P. Atcheson}
\affiliation{CRANN, AMBER and School of Physics, Trinity College Dublin,
Ireland}
\author{Roberto E. Troncoso}
\author{Arne Brataas}
\affiliation{Center for Quantum Spintronics, Department of Physics, Norwegian
University of Science and Technology, NO-7491 Trondheim, Norway}
\author{J.M.D.\,Coey}
\author{Plamen Stamenov}
\author{Karsten Rode}
\email[Corresponding author:]{rodek@tcd.ie}
\affiliation{CRANN, AMBER and School of Physics, Trinity College Dublin,
Ireland}

\maketitle

\subsection{Sample preparation}
\label{ssec:sampleprep}
The samples used in this study are thin films of \mrx\ with $x=0.7$ (\mrg) grown on
single-crystal MgO $\left( 001 \right)$ substrates by magnetron co-sputtering
from  Mn$_2$Ga and Ru targets in a `Shamrock' deposition cluster tool at a
substrate temperature $T_\text{s} \approx \SI{350}{\degreeCelsius}$. The Ru
concentration $x$ was varied by changing the power to the Mn$_2$Ga target while
keeping that of Ru constant. After deposition of the active \mrg\ layer, the
samples were covered, at room temperature, with $\sim\SI{2}{\nano\metre}$ of
AlO$_x$ to prevent oxidation. After deposition, the thin films were
patterned into Hall bars using standard UV lithography techniques and Ar-ion
dry etching. Contacts were made using Ti/Au patterned by lift-off. The samples
had the long axis of the Hall bars parallel to MgO
$\left[ 110 \right]$, which is parallel to \mrg\ $\left[ 100 \right]$ or
$\left[ 010 \right]$. We verified the crystal structure of every sample using
X-ray diffraction collected using a Bruker D8 Discovery instrument with a
primary double-bounce Ge $\left[ 110 \right]$ monochromator, as well as their
thickness and interface roughness via small-angle X-ray reflectivity in the
same diffractometer. Film thicknesses were \SI{30}{\nano\metre}, with a surface roughness
$< \SI{0.7}{\nano\metre}$.

\subsection{Crystallography}
\label{ssec:crystalsymmetry}
\mrg\ crystallises in the cubic space group $F\bar{4}3m$ with the symmetry
reduced to $I\bar{4}m2$ due to the tetragonal distortion induced by biaxial
substrate strain. In this tetragonal representation the current-carrying Mn
sublattice occupies the Wyckoff $2c$ position with point-group
symmetry $\bar{4}m2$, corresponding to the $4c$ position in $F\bar{4}3m$ with
point group symmetry $\bar{4}3m$. The tetragonal representation has its basal
plane rotated by \SI{45}{\degree} with respect to the cubic one, hence in order
to retain the coordinate system of the cubic representation, we consider the
point symmetry of Mn in the tetragonal representation to be $\bar{4}2m$. For
the remainder of this text, we use the cubic representation of \mrg\ when
referring to directions in both the direct and reciprocal space.

The growth mode of the sputtered films is coalescing-island-like, and a high degree of in-plane order is
achieved so that the epitaxial relation can be written $\text{MRG} \left[ 001
\right] \parallel \text{MgO} \left[ 001 \right]$ and $\text{MRG} \left[ 110
\right] \parallel \text{MgO} \left[ 100 \right]$. Due to the
growth technique and mode, there is no difference between $\text{MRG} \left[ 100
\right]$ and $\left[ 010 \right]$, and likewise for the other in-plane
directions. In the following we will apply an electric bias current along the
different crystallographic directions of \mrg, for example $\left[ 100 \right]$.
Due to the twinning, half the crystallites will see a current along
$\left[ 100 \right]$ and the other half along $\left[ 010 \right]$.

\subsection{Form of the current-induced effective field}
\label{ssec:sot-field}
The form of the effective field for all the point groups that lack spatial
inversion symmetry, a requirement for intrinsic spin-orbit torque (SOT), has been
tabulated by \citet{zelezny2017}, and discussed specifically for \mrg\  by
\citet{troncoso2018antiferromagnetic}.
For the point group $\bar{4}2m$, in the exchange approximation and omitting the
dot product with the dimensionless unit current density vector
\begin{equation}
  \label{eq:sottetragonal}
  h_{\text{SOT}}^{\text{tet}} = x_{fl}
  \begin{pmatrix}
    1 & 0 & 0 \\
    0 & -1 & 0 \\
    0 & 0 & 0
  \end{pmatrix}  +
  \begin{pmatrix}
    0 & m_z x_3 & m_y x_2 \\
    m_z x_3 & 0 & m_x x_2 \\
    m_y x_1 & m_x x_1 & 0
  \end{pmatrix}
\end{equation}
where $h_{\text{SOT}}$ is the SOT effective field, and $x_i$ are coefficients
that relate the magnitude of the current density to the magnitude of the field. $m_i$ are the
components of the net magnetisation unit vector. For simplicity we give all
fields in \si{\tesla} and the unit of $x_i$ and $h_{\text{SOT}}$ is therefore
\si{\tesla\per\ampere\metre\squared}. In line with the material properties
given above, current-carrying Mn in \mrg\ can be considered to be in an almost
cubic environment with point group symmetry $\bar{4}3m$. The effective field is
\begin{equation}
  h_{\text{SOT}}^{\text{cubic}} = x_{dl}
  \begin{pmatrix}
    0 & m_z & m_y \\
    m_z & 0 & m_x \\
    m_y & m_x & 0
  \end{pmatrix}
\end{equation}
We take advantage of the near-cubic nature of \mrg\ to ignore the second term
in~\eqref{eq:sottetragonal} and reduce the number of
unknown parameters; the symmetry of the effective field should
simultaneously reflect the symmetry of $\bar{4}3m$ and $\bar{4}2m$:
\begin{equation}
  h_{\text{SOT}} = x_{fl}
  \begin{pmatrix}
    1 & 0 & 0 \\
    0 & -1 & 0 \\
    0 & 0 & 0
  \end{pmatrix}  + x_{dl}
  \begin{pmatrix}
    0 & m_z & m_y \\
    m_z & 0 & m_x \\
    m_y & m_x & 0
  \end{pmatrix}
  \label{eq:sotmatrix}
\end{equation}
Here we recognise two terms, one reactive (field-like) $x_{fl}$ that does not
depend on the magnetisation and one dissipative (damping-like) $x_{dl}$ that
does.

In this approximation we can describe the SOT in \mrg\ through just two material
parameters, the coefficients $x_{fl}$ and $x_{dl}$, reducing the number of free parameters from
eight (four per sublattice) to two, greatly improving the significance of fits (and analysis) while disregarding minute differences in the spatial anisotropy of the
tetragonal representation.

The above does not imply that SOT cannot alter the angle between the two sublattices,
both statically and dynamically. In the real physical system, the
two lattices have different presence at $E_{\text{F}}$, and they experience
different torques due to the conduction electrons. As they correspondingly have
different anisotropies and exchange stiffness, SOT is bound to influence both
the zero- and finite-frequency antiferromagnetic magnons. How much energy is
pumped into the in-phase and anti-phase modes would strongly depend on the
degree of compensation. Excitation at the high-frequency antiferromagnetic
mode, will likely only be efficient in a region close to compensation.

\subsection{Relating the effective field to a change in magnetisation}
\label{ssec:deltatheta}
The magnetic part of the free energy in a tetragonal system is 
written, to second order in anisotropy as
\begin{align}
  E &= K_1 \sin^2{\left( \theta \right)} + K_2 \sin^4{\left( \theta \right)} + K_2^{'}
  \sin^4{\left( \theta \right)}
  \cos{\left( 4 \phi_M \right)} \notag \\
  &- \mu_0 \vec{H}\left( \theta_H, \phi_H \right) \cdot \vec{M}
  \left( \theta, \phi_M \right)
\end{align}
where $\theta$ and $\phi_M$ are the polar and azimuthal angles of the
magnetisation. $K_2^{'}$ is usually much weaker than $K_{1,2}$ and we assume
the in-plane component of the magnetisation is aligned with the applied field
$\mu_0 \vec{H}$, $\phi_H = \phi_M$.
The equilibrium position of the magnetisation can then be derived from the
stability condition 
\begin{equation}
  F = \pdv{E}{\theta} = 0
  \label{eq:stability}
\end{equation}
although equation \eqref{eq:stability} cannot be directly solved for $\theta$. The aim,
however, is to relate a small effective field $\mathbf{h_{\text{SOT}}}$ to a (small)
change in $\theta$. We write $\mathbf{H} = \mathbf{H_0} +
\mathbf{h_{\text{SOT}}}$ and $\theta = \theta_0
+ \Delta\theta$, and note that $F(\theta_0,\mathbf{H_0}) = 0$ (the equilibrium
position), as well as $\pdv{F}{H_i}{H_j} = 0$ because $E$ is linear in
$\mathbf{H}$.

We now expand $F$ to second order in $\mathbf{h_{\text{SOT}}}$, which leads to a
quadratic equation. Realising that $\pdv{F}{\theta} \gg \mathbf{h_{\text{SOT}}}$, we
expand the square root portion of the solution to that quadratic equation, and
select the solution that ensures $F = 0$ when $\mathbf{h_{\text{SOT}}} = 0$.

After some straightforward but tedious algebra we find:
\begin{align}
  - \Delta\theta =&  \left(\pdv{F}{\theta}\right)^{-1} \left( \mathbf{v_1}
    \cdot \mathbf{h_{\text{SOT}}}
  \right) \notag \\
  +& \frac{\left( \pdv{F}{\theta} \right)^{-2}}{2} \left( \mathbf{v_1} \cdot
  \mathbf{h_{\text{SOT}}} \right) \left( \mathbf{v_2} \cdot
  \mathbf{h_{\text{SOT}}} \right)
\end{align}
with 
\begin{equation}
  \mathbf{v_1} = \pdv{F}{\mathbf{H}} = 
  \begin{pmatrix}
    \pdv{F}{H_x} \\ \pdv{F}{H_y} \\ \pdv{F}{H_z}
  \end{pmatrix}
\end{equation}
and
\begin{equation}
  \mathbf{v_2} = 
  \begin{pmatrix}
    2 \pdv{F}{\theta}{H_x} - \pdv{F}{\theta}{\theta} \pdv{F}{H_x} 
    \frac{1}{\pdv{F}{\theta}} \\
    2 \pdv{F}{\theta}{H_y} - \pdv{F}{\theta}{\theta} \pdv{F}{H_y}
    \frac{1}{\pdv{F}{\theta}} \\
    2 \pdv{F}{\theta}{H_z} - \pdv{F}{\theta}{\theta} \pdv{F}{H_z}
    \frac{1}{\pdv{F}{\theta}}
  \end{pmatrix}
\end{equation}
which we rewrite as
\begin{equation}
  \Delta\theta = \Delta\theta_1 j + \Delta\theta_2 j^2
\end{equation}
where the units of $\Delta\theta_1$ and $\Delta\theta_2$ are
\si{\radian\per\ampere\metre\squared} and
\si{\radian\per\ampere\squared\metre\tothe{4}}, respectively, because
$h_{\text{SOT}}$ is linear in current density.

\subsection{Harmonic Hall}
\label{ssec:harmonichall}
The longitudinal and transverse voltages were measured using lock-in detection at
the first, second and third harmonic, at a low excitation
frequency, typically $f_{\text{AC}} \sim \SI{1}{\kilo\hertz}$. The excitation
frequency is far from the frequency of magnetisation or thermal dynamics, but
provides a convenient means to separate effects that are odd or even functions
of the bias current.

Materials that exhibit high spin-orbit coupling are often quite resistive and
are therefore prone to heating due to the relatively high current densities
required to observe SOT\@. In addition, the high spin-orbit coupling also leads
to strong magneto-thermal effects, in particular normal and anomalous Nernst
effects, which are impossible to separate from SOT on the second harmonic using
only AC bias current. We therefore prefer to use either the difference between
AC + DC and AC - DC on the second harmonic, as the Nernst effect is quadratic
in DC current while the SOT is linear, \emph{or} record data on the third
harmonic where the Nernst effet is not present. Homogeneous heating, $\Delta
T$, of the sample is present in both cases however.

We therefore write $\rho = \sigma^{-1}$, $\sigma_{xx} = \sigma_{xx}^0 (1 +
\alpha)$, $\sigma_{xy} = \sigma_{xy}^0 (1 + \beta)$ with $\alpha =
\pdv{\sigma_{xx}}{T} \Delta T / \sigma_{xx}^0$ and $\beta =
\pdv{\sigma_{xy}}{T} \Delta T / \sigma_{xy}^0$.
For \mrg\ the ratio $\sigma_{xy}/\sigma_{xx} \sim \SI{10}{\percent}$ so that
$\rho_{xx} \approx \sigma_{xx}^{-1}$ and $\rho_{xy} \approx \sigma_{xy} /
\sigma_{xx}^2$ with an error of $\approx \SI{1}{\percent}$.

We now include the relations above, with $\cos{\left( \theta_0 + \Delta\theta
\right)}$ accounting for the $m_z$ dependence of the transverse voltage, to find
\begin{equation}
  \rho_{xy} = \frac{\sigma_{xy}^0(1 + \beta)}{\left(\sigma_{xx}^0 (1 + \alpha)\right)^2}
  \cos{\left( \theta_0 + \Delta\theta \right)}
\end{equation}
which we expand to first order in $\alpha$ and $\beta$, and second order in
$\Delta\theta$. $\Delta T$ is due to resistive heating, $\Delta T = \chi I^2$,
where $\chi$ characterises the conversion from current to heat and is related
to the resistivity of the sample. Likewise, $\Delta\theta$ is replaced by its
explicit dependence on current density $\Delta\theta = \Delta\theta_1 j +
\Delta\theta_2 j^2$. $I$ is the bias current, while $j$ is the bias current
density. Obtaining the voltage response on each harmonic of the bias current is
then a matter of expanding the powers of the sinusoidal current $I = I_{\text{ac}}
\sin{ \omega t} + I_{\text{dc}}$, and projecting on $\sin{\omega t}$ (first
harmonic), $\cos{2 \omega t}$ (second harmonic) and $\sin{3 \omega t}$ (third
harmonic).

Finally, we obtain
\begin{align}
  \left( V_{xy}^{2 \omega}(I_{\text{ac}}, I_{\text{dc}}) - V_{xy}^{2
  \omega}(I_{\text{ac}}, -I_{\text{dc}})\right) /  I_{\text{ac}}^2 I_{\text{dc}} =  \notag \\
  \frac{6 B_z  \alpha \chi}{t (\sigma_{xx}^0)^3} \sigma_{\text{oh}} 
  + \frac{3 l \alpha \chi}{t w (\sigma_{xx}^0)^2}
  + \frac{3 \Delta\theta_1^2 \sigma_{xy}^0 \cos \theta_0}{2 t^3 w^2
  (\sigma_{xx}^0)^2} \notag \\
  - \frac{\beta \chi \cos{\theta_0}}{t (\sigma_{xx}^0)^2}
  + \frac{6 \alpha \sigma_{xy}^0 \chi \cos{\theta_0}}{t (\sigma_{xx}^0)^3}
  + \frac{3 \Delta\theta_2 \sigma_{xy}^0 \sin{\theta_0}}{t^3 w^2
  (\sigma_{xx}^0)^2}
  \label{eq:2w}
\end{align}
and
\begin{align}
  \left(V_{xy}^{3 \omega}(I_{\text{ac}},0)\right) / I_{\text{ac}}^3 = \notag \\
  \frac{B_z \alpha \sigma_{\text{oh}} \chi}{2 t (\sigma_{xx}^0)^3}
  + \frac{l \alpha \chi}{4 t w (\sigma_{xx}^0)^2}
  + \frac{\Delta\theta_1^2 \sigma_{xy}^0 \cos{\theta_0}}{8 t^3 w^2
  (\sigma_{xx}^0)^2} \notag \\
  - \frac{\beta \chi \cos{\theta_0}}{4 t (\sigma_{xx}^0)^2}
  + \frac{\alpha \sigma_{xy}^0 \chi \cos{\theta_0}}{2 t (\sigma_{xx}^0)^3}
  + \frac{\Delta\theta_2 \sigma_{xy}^0 \sin{\theta_0}}{4 t^3 w^2 (\sigma_{xx}^0)^2}
  \label{eq:3w}
\end{align}
where $l$ is the inevitable small offset of one Hall contact with respect to
the other due to misalignment during lithography,
$w$, $t$ are the Hall bar width and thickness, and we have included the normal
Hall conductivity, $\sigma_{\text{oh}}$. Similar equations can be written out
for the harmonics of the longitudinal ($V_{xx}^{n \omega}$, $n = 1, 2, 3$) signals.

It is useful to relate the above equations~\eqref{eq:2w} and \eqref{eq:3w} to the
system we are measuring. As can be seen from Eq.~\eqref{eq:sotmatrix}, in the
case of statistical twinning of the crystal of \SI{90}{\degree} around the
$c$ axis, most of the SOT terms cancel each other (with the exception of a
dissipative term that depends on the in-plane magnetisation). The torque still
exists however, inside each crystallographic domain, and the correct way of
analysing the data therefore, in the case of terms in
$\Delta\theta^2$, is to evaluate the sum of the squares rather than the square of
the sums. We therefore expect to see the signature (in the $\phi$ dependence)
of both the reactive and the dissipative components of the torque via the term
that depends on $\Delta\theta_1^2$.

The extracted parameters for the field- and damping-like contributions to the
SOT are then obtained by fitting the model to the experimental data, using
values for $\sigma$, $\alpha$, $\beta$ and anisotropy constants determined by
experiment.

\subsection{Experimental Hall measurements}
\label{ssec:multimag}
We recorded harmonic Hall data in two different experimental configurations.
 
First, for the data collected in the low-current-density regime, a
`Multimag' permanent-magnet-based magnetotransport setup was used. The Multimag is a permanent magnet variable flux source consisting
of two concentric Halbach cylinders that each provide a field of $\mu_0 H
\approx \SI{1}{\tesla}$, mounted on ball-bearings with their angular postion
controlled by servo-motors and optical encoder feedback. With this arrangement
 a field of up to $\SI{2}{\tesla}$ can be created in any direction
$\phi_H$ in the horizontal plane, which is very convenient for magnetotransport measurements as a function of angle. After saturation in an out-of-plane field we
recorded the frist, second and third harmonic of the transverse voltage as a
function of field magnitude and azimuthal angle $\phi_H$. All measurements were
done at room temperature without temperature control. We also recorded
out-of-plane loops to calibrate the relation between $V_{xy}$ and $m_z$. The
lock-in detection was done with two Zürich Instruments MFLI lock-in amplifiers
which allowed recording $V_{xx}$ and $V_{xy}$ at the three harmonics
simultaneously.

Second, we collected data using a Quantum Design `Physical Properties
Measurement System' with a maximum field of $\pm \SI{14}{\tesla}$, this time
using a battery of Standford Research 830 lock-in amplifiers. The different
amplifiers were cross calibrated to ensure the magnitudes at each harmonic are
related in a physically meaningful way, and the phase-contribution from the
measurement apparatus determined by fitting the frequency-dependence of the
signal recorded on an ohmic resistor. Again, we recorded out-of-plane loops
to determine the relation between $V_{xy}$ and $m_z$. We also recorded in-plane
loops from \SIrange{-14}{14}{\tesla}, which allows us to determine a relation
between the observed $m_z$ and the effective field acting on the magnetisation.
In detail, both the out-of-plane and the in-plane loops were modelled using a classical micromagnetic
approach (Preisach model), based on statistical distribution of an ensemble of hysterons that
switch fully at a discrete applied field. In-plane, $\phi_H$ -scan, for varying current densities, has been measured by rotating the sample around $e_z$ axis in the presence of applied field of magnitude \SIlist{0,2; 0.4; 0.6}{\tesla}. Special care has been taken to maintain the sample temperature ($ T=\SI{310}{\kelvin} $) while altering the current densities.

In the main text we determine the transfer function relating $I$ to an
effective field by analysing the variation of $V_{xy}$ with current density. In
the first part we consider this relation to be linear, and can
therefore use the symmetry relations given above to determine the full SOT
tensor. In the second part of the main text this linear relation breaks down
due to the high effective field created by the SOT\@. We therefore (in both
cases) take into account the linear contributions of the SOT on each harmonic,
and by Fourier analysis find that the first order contribution
to $V_{xy}$ can be isolated by calculating $V_{xy}^{ \omega} + 3
V_{xy}^{3 \omega}$. Any non-linearity observed in \figurename~4 of the main
text
therefore reflects non-linearities in the transfer function itself (due to SOT), as we have
isolated the contribution that depends on $\phi_M$.

\bibliography{bibliography}